\documentclass[aps,prb,twocolumn,showpacs,floatfix,longbibliography,10pt]{revtex4-1}
\usepackage{graphicx,amsfonts,amssymb,amsmath}
\usepackage{textcomp}
\usepackage{esint}
\usepackage[colorlinks,linktocpage,bookmarks=false,citecolor=blue,linkcolor=red,urlcolor=blue]{hyperref}
\usepackage[utf8]{inputenc}
\usepackage[T1]{fontenc}
\usepackage[dvipsnames]{xcolor}
\usepackage[english]{babel}
\usepackage{afterpage}

\allowdisplaybreaks

\newlength{\textlarg}

\def\beq{\begin{equation}}
\def\eeq{\end{equation}}
\def\bleq{\begin{eqnarray}}
\def\eleq{\end{eqnarray}} 
\def\bfig{\begin{figure}}
\def\efig{\end{figure}}
\def\bline{\begin{multline}}
\def\eline{\end{multline}}
\def\bremark{\begin{quotation} \noindent \small }
\def\eremark{\end{quotation}}

\newcommand{\Tr}{{\rm Tr}}

\newcommand{\mean}[1]{\langle #1 \rangle}

\def\eps{\epsilon}
 
\def\phibf{\boldsymbol{\phi}}
\def\varphibf{\boldsymbol{\varphi}}

\def\Lamb{\Lambda}
\def\sig{\sigma}
\def\Sig{\Sigma}

\def\half{\frac{1}{2}}

\def\quarter{\frac{1}{4}}

\def\p{{\bf p}} 
\def\q{{\bf q}}
\def\r{{\bf r}}

\def\v{{\bf v}}
\def\x{{\bf x}}

\def\A{{\bf A}}

\def\J{{\bf J}}

\def\w{\omega}
\def\wn{\omega_n}

\def\dmu{{\partial_\mu}}

\def\dk{\partial_k}

\def\dtau{{\partial_\tau}}

\def\inttau{\int_0^\beta d\tau}

\def\calD{{\cal D}}

\def\calO{{\cal O}}
  
\def\calR{{\cal R}}

\def\calZ{{\cal Z}}

\def\twn{{\tilde\omega_n}}

\graphicspath{{./figures/}}

\def\Jmua{J^a_\mu} 
 
\def\jmua{j^a_\mu}  
 
\def\jnub{j^b_\nu} 
\def\Amu{A_\mu} 
 
\def\Amua{A^a_\mu}

\def\Kmunuab{K^{ab}_{\mu\nu}}

\def\Pimunuab{\Pi^{ab}_{\mu\nu}}

\def\sigmunuab{\sig^{ab}_{\mu\nu}}
\def\Sigmunuab{\Sig^{ab}_{\mu\nu}}

\def\Cdis{C_{\rm dis}} 
\def\Lord{L_{\rm ord}}

\def\sigA{\sig_{\rm A}}
\def\sigB{\sig_{\rm B}}
\def\sigAk{\sig_{{\rm A},k}}
\def\sigBk{\sig_{{\rm B},k}}
\def\sigQ{\sig^{}_Q}

\begin{document}

\title{Superuniversal transport near a $(2+1)$-dimensional quantum critical point}

\author{F. Rose}
\author{N. Dupuis}
\affiliation{Laboratoire de Physique Th\'eorique de la Mati\`ere Condens\'ee,
CNRS UMR 7600, UPMC-Sorbonne Universit\'es, 4 Place Jussieu, 
75252 Paris Cedex 05, France}

\date{September 5, 2017)
} 

\begin{abstract}
We compute the zero-temperature conductivity in the two-dimensional quantum O($N$) model using a nonperturbative functional renormalization-group approach. At the quantum critical point we find a universal conductivity $\sigma^*/\sigQ$ (with $\sigQ=q^2/h$ the quantum of conductance and $q$ the charge) in reasonable quantitative agreement with quantum Monte Carlo simulations and conformal bootstrap results. In the ordered phase the conductivity tensor is defined, when $N\geq 3$, by two independent elements, $\sigma_{\rm A}(\w)$ and $\sigma_{\rm B}(\w)$, respectively associated with SO($N$) rotations which do and do not change the direction of the order parameter. Whereas $\sigA(\w\to 0)$ corresponds to the response of a superfluid (or perfect inductance), the numerical solution of the flow equations shows that $\lim_{\w\to 0}\sigB(\w)/\sigQ=\sigB^*/\sigQ$ is a superuniversal (i.e. $N$-independent) constant. These numerical results, as well as the known exact value $\sigB^*/\sigQ=\pi/8$ in the large-$N$ limit, allow us to conjecture that $\sigB^*/\sigQ=\pi/8$ holds for all values of $N$, a result that can be understood as a consequence of gauge invariance and asymptotic freedom of the Goldstone bosons in the low-energy limit.   
\end{abstract}

\maketitle

{\bf Introduction.}
Understanding the physical properties of a system near a quantum phase transition constitutes an important problem in condensed-matter physics. This is particularly true of transport properties since strong fluctuations near the quantum critical point (QCP) often lead to the absence of well-defined quasi-particles and the breakdown of perturbation many-body theory.\cite{Sachdev_book}

In this Rapid Communication, we discuss the zero-temperature coherent transport near a relativistic $(2+1)$-dimensional QCP with an O($N$)-symmetric order parameter. We use a nonperturbative functional renormalization-group (NPRG) approach to compute the frequency-dependent conductivity in the quantum O($N$) model. The latter describes many condensed-matter systems with a relativistic effective low-energy dynamics: quantum antiferromagnets, bosons in optical lattices, Josephson junctions, etc. At the QCP we find a universal conductivity\cite{Fisher90,not6} $\sig^*/\sigQ$ (with $\sigQ=q^2/h$ the quantum of conductance and $q$ the charge) in reasonable quantitative agreement with quantum Monte Carlo simulations\cite{Witczak14,Chen14,Gazit13a,Katz14,Gazit14} and conformal bootstrap results.\cite{Kos15} Our main result concerns the broken-symmetry phase, where the conductivity tensor has two independent elements $\sigA$ and $\sigB$ when $N\geq 3$, respectively associated with SO($N$) rotations which do and do not change the direction of the order parameter. Whereas $\sigA(\w\to 0)$ corresponds to the response of a superfluid (or perfect inductance), the numerical solution of the NPRG equations together with the exact large-$N$ result leads us to conjecture that $\sigB(\w\to 0)/\sigQ$ takes the superuniversal (i.e. $N$-independent) value $\sigB^*/\sigQ=\pi/8$.  We argue that this result is a consequence of gauge (rotation) invariance and asymptotic freedom in the infrared, i.e. the fact that Goldstone bosons become effectively noninteracting in the low-energy limit. 

This conjecture has been anticipated in Ref.~\onlinecite{Rose17} using an approximate solution of the NPRG equations based on a derivative expansion of the scale-dependent effective action. However, because of infrared singularities which invalidate the derivative expansion at low energy, we could not obtain definite values for $\sig^*/\sigQ$ and $\sigB^*/\sigQ$. Here we report results obtained from a different approximation of the NPRG equations which does not suffer from these limitations. 

While universality is a generic consequence of the proximity of the QCP, universal quantities (e.g. critical exponents or scaling functions) in general depend on $N$. To our knowledge there are very few exceptions. The critical energy densities of O($N$) models on a $d$-dimensional lattice with long-range interactions are known to be all equal to the one of the Ising model.\cite{Campa03} The same is true for all O($N$) models on a one-dimensional lattice with nearest-neighbor interactions. It has been conjectured\cite{Casetti11} that this superuniversality should hold for all $d$-dimensional O($N$) models but a firm numerical confirmation has not been provided so far.\cite{Nerattini14}

{\bf Quantum O($N$) model and NPRG approach.} 
The two-dimensional quantum O($N$) model is defined by the Euclidean action
\begin{equation}
S = \int_\x \biggl\lbrace \half \sum_{\mu=0,x,y}(\dmu\varphibf)^2  
+ \frac{r_0}{2} \varphibf^2 + \frac{u_0}{4!N} {(\varphibf^2)}^2  \biggr\rbrace ,
\label{action1} 
\end{equation}
where we use the notation $\x=(\r,\tau)$, $\int_\x=\inttau \int d^2r$ and $\partial_0=\dtau$. $\varphibf(\x)$ is an $N$-component real field,  $\r$ a two-dimensional coordinate, $\tau\in [0,\beta]$ an imaginary time, and $\beta=1/T$ the inverse temperature (we set $\hbar=k_B=1$). $r_0$ and $u_0$ are temperature-independent coupling constants and the (bare) velocity of the $\varphibf$ field has been set to unity. The model is regularized by an ultraviolet cutoff $\Lamb$. Assuming $u_0$ fixed, there is a quantum phase transition between a disordered phase ($r_0>r_{0c}$) and an ordered phase ($r_0<r_{0c}$) where the O($N$) symmetry is spontaneously broken. The QCP at $r_0=r_{0c}$ is in the universality class of the three-dimensional classical O($N$) model and the phase transition is governed by the three-dimensional Wilson-Fisher fixed point. 

In the following we consider only the zero-temperature limit where the two-dimensional quantum model is equivalent to the three-dimensional classical model. We thus identify $\tau$ with a third spatial dimension so that $\x=(\r,\tau)\equiv (x,y,z)$. A correlation function $\chi(p_x,p_y,p_z)$ computed in the classical model then corresponds to the correlation function $\chi(p_x,p_y,i\wn)$ of the quantum model, with $\wn\equiv p_z$ a bosonic Matsubara frequency,\footnote{At zero temperature, the bosonic Matsubara frequency $\wn=2n\pi T$ ($n$ integer) becomes a continuous variable.} and yields the retarded dynamical correlation function $\chi^R(p_x,p_y,\w)$ after analytical continuation $i\wn\to\w+i0^+$.

The O($N$) symmetry of the action~(\ref{action1}) implies the conservation of the total angular momentum and the existence of a conserved current. To compute the associated conductivity, we include in the model an external non-Abelian gauge field $A_\mu=A_\mu^a T^a$ (with an implicit sum over repeated discrete indices), where $\{T^a\}$ denotes a set of SO($N$) generators (made of $N(N-1)/2$ linearly independent skew-symmetric matrices). This amounts to replacing the derivative $\dmu$ in Eq.~(\ref{action1}) by the covariant derivative $D_\mu=\dmu-q A_\mu$ (we set the charge $q$ equal to unity in the following and restore it, as well as $\hbar$, whenever necessary). This makes the action~(\ref{action1}) invariant in the local gauge transformation $\varphibf'=O\varphibf$ and $\Amu'= O \Amu O^T +(\dmu O)O^T$ where $O$ is a space-dependent SO($N$) rotation. 
The current density $\Jmua(\x) = - \delta S/\delta \Amua(\x)$ is then expressed as~\cite{Rose17} 
\beq
\Jmua =  \jmua -\Amu\varphibf \cdot T^a \varphibf ,  \quad
\jmua = \dmu \varphibf \cdot T^a \varphibf ,
\eeq
where $\jmua$ denotes the ``paramagnetic'' part. For $N=2$, there is a single generator $T$, which can be
chosen as minus the antisymmetric tensor $\eps_{ij}$,\cite{Rose17} and we recover the standard expression $j_\mu=-i[\psi^* \dmu\psi-(\dmu\psi^*)\psi]$ of the current density of bosons described by a complex field $\psi=(\varphi_1+i\varphi_2)/\sqrt{2}$. 
For $N=3$, there are three generators $-iS^1$, $-iS^2$ and $-iS^3$ related to spin-one matrices $S^i$. One then finds $j_\mu^i=-\eps_{ijk}(\dmu\varphi_j)\varphi_k$ ($\eps_{ijk}$ is the antisymmetric tensor) in agreement with the continuum limit of spin currents defined in lattice models.\cite{[{See, e.g., }]Rueckriegel17} 

The frequency-dependent conductivity of the quantum model is defined as the linear response to the gauge field, i.e. 
\beq
\sigmunuab(\w) = - \frac{i}{(\w+i0^+)} \Kmunuab{}^R(\w) .
\label{sig1}
\eeq
$K^R$ is the retarded part of the correlation function 
\beq
\Kmunuab(i\wn)=\Pimunuab(i\wn)-\delta_{\mu\nu} \mean{T^a\varphibf\cdot T^b\varphibf} ,
\eeq
where we have set the momentum to zero and $\Pimunuab=\mean{\jmua\jnub}$ is the paramagnetic current-current correlation function. The conductivity having a vanishing scaling dimension in two space dimensions, it satisfies\cite{Fisher90,Damle97}
\beq
\sigmunuab(\w) = \sigQ \Sigmunuab{}_\pm \left( \frac{\w+i0^+}{\Delta} \right) , 
\eeq 
where $\Sigmunuab{}_\pm$ is a universal scaling function (the index $+/-$ refers to the disordered/ordered phase), and $\Delta$ a characteristic zero-temperature energy scale which measures the distance to the QCP. In the disordered phase, we take $\Delta$ to be equal to the excitation gap. In the ordered phase, we choose $\Delta$ to be given by the excitation gap in the disordered phase at the point located symmetrically with respect to the QCP (i.e. corresponding to the same value of $|r_0-r_{0c}|$). The conductivity tensor is diagonal in the disordered phase so that a single scaling function $\Sigma_+$ has to be considered. In the ordered phase, it has only two independent elements, $\sigA$ and $\sigB$, respectively associated with SO($N$) rotations which do and do not change the direction of the order parameter.\cite{Rose17} [For $N=2$ there is only one generator and the conductivity is diagonal also in the ordered phase.]  

The strategy of the NPRG approach is to build a family of models indexed by a momentum scale $k$ such that fluctuations are smoothly taken into account as $k$ is lowered from the microscopic scale $\Lambda$ down to 0.\cite{Berges02,Delamotte12,Kopietz_book} This is achieved by adding to the action $S[\varphibf,\A]$ the gauge-invariant infrared regulator term~\cite{Rose17} 
\beq
\Delta S_k[\varphibf,\A] = \half \int_\x \varphibf \cdot R_k\bigl(-D^2\bigr) \varphibf , 
\label{DeltaSk}
\eeq
where $D^2=D_\mu D_\mu$. The partition function 
\begin{equation}
\calZ_k[\J,\A] = \int\calD[\varphibf]\, e^{-S[\varphibf,\A]-\Delta S_k[\varphibf,\A]+\int_\x \J\cdot\varphibf}  
\end{equation} 
is now $k$ dependent. Here $\J$ is an external source which couples linearly to the $\varphibf$ field. The order parameter $\phibf_k[\x;\J,\A]=\delta\ln Z_k[\J,\A]/\delta\J(\x)=\mean{\varphibf(\x)}$ is a functional of both $\J$ and $\A$. The scale-dependent effective action 
\begin{equation}
\Gamma_k[\phibf,\A] = -\ln \calZ_k[\J,\A] + \int_\x \J\cdot\phibf -\Delta S_k[\phibf,\A] 
\end{equation}
is defined as a (slightly modified) Legendre transform of $-\ln \calZ_k[\J,\A]$, where the linear source $\J\equiv\J_k[\phibf,\A]$ is now considered as a functional of $\phibf$ and $\A$. Assuming that fluctuations are completely frozen by the $\Delta S_k$ term when $k=\Lambda$, $\Gamma_\Lambda[\phibf,\A]=S[\phibf,\A]$. On the other hand the effective action of the original model, defined by the action $S[\varphibf,\A]$, is given by $\Gamma_{k=0}$ provided that $R_{k=0}$ vanishes. 
The variation of the effective action with $k$ is given by Wetterich's equation\cite{Wetterich93} 
\begin{equation}
\dk \Gamma_k[\phibf,\A] = \half \Tr \Bigl\{ \dk \calR_k[\A] \bigl(\Gamma_k^{(2,0)}[\phibf,\A] + \calR_k[\A] \bigr)^{-1} \Bigr\} ,
\label{Weteq}
\end{equation}
where $\Gamma_k^{(2,0)}[\phibf,\A]$ and $\calR_k[\A]$ denote the second-order functional derivative with respect to $\phibf$ of $\Gamma_k[\phibf,\A]$ and $\Delta S_k[\phibf,\A]$, respectively. In Fourier space, the trace involves a sum over momenta as well as the O($N$) index of the $\phibf$ field. The conductivity of the quantum model is calculated using 
\beq
\Kmunuab(\p,i\wn) = -\Gamma_{k=0,\mu\nu}^{(0,2)ab}(\p,i\wn) ,
\label{KGam02}
\eeq
where $\Gamma_k^{(0,2)}$ is the second-order functional derivative of $\Gamma_k[\phibf,\A]$ with respect to $\A$, 
evaluated for $\A=0$ and in the uniform time-independent field configuration $\phibf$ which minimizes the effective action $\Gamma_{k=0}[\phibf,\A=0]$.\cite{Rose17} 

To solve Eq.~(\ref{Weteq}) we consider the following gauge-invariant ansatz
\begin{multline}
\Gamma_k[\phibf,\A] = \int_\x \biggl\lbrace U_k(\rho) + \half D_\mu\phibf \cdot Z_k(-D^2)  D_\mu\phibf \\ + \quarter (\partial_\mu \rho) Y_k(-\partial^2) (\partial_\mu \rho)  
+ \dfrac{1}{4} F_{\mu \nu}^a X_{1,k}(-D^2) F_{\mu \nu}^a  \\ + \dfrac{1}{4}  F_{\mu \nu}^a T^a\phibf \cdot X_{2,k}(-D^2) F_{\mu \nu}^b T^b\phibf \biggr\rbrace ,
\label{lpapp} 
\end{multline}
which, in addition to the effective potential $U_k(\rho=\phibf^2/2)$, involves four functions of momentum: $Z_k(\q^2)$, $Y_k(\q^2)$, $X_{1,k}(\q^2)$ and $X_{2,k}(\q^2)$. This approximation, which we dub LPA$''$, has been used in the past to compute the critical indices and the momentum dependence of correlation functions in the O($N$) model in the absence of the gauge field.\cite{Hasselmann12, Canet10,*Canet11,*Canet11a,*Canet16} For $Z_k(\q^2)\equiv Z_k$ and $Y_k(\q^2)=0$ it reduces to the LPA$'$, an improvement of the local potential approximation (LPA) which includes a field-renormalization factor $Z_k$.\cite{Berges02,Delamotte12} We denote by $\rho_{0,k}$ the value of $\rho$ at the minimum of the effective potential. Spontaneous breaking of the O($N$) symmetry is characterized by a nonvanishing value of $\rho_{0,k}$ for $k\to 0$.

From Eqs.~(\ref{lpapp}) and (\ref{Weteq}) we obtain RG equations for the functions $U_k$, $Z_k$, $Y_k$, $X_{1,k}$ and $X_{2,k}$, and in turn for the vertex 
\begin{multline}
\Gamma_{k,\mu\nu}^{(0,2)ab}(\p=0,i\wn) = \delta_{\mu\nu} \delta_{ab} \wn^2 X_{1,k}(\wn^2) \\  +
\delta_{\mu\nu} (T^a\phibf)\cdot (T^b\phibf) [ Z_k(\wn^2) + \wn^2 X_{2,k}(\wn^2) ] ,
\end{multline}
which determines the conductivity [Eqs.~(\ref{sig1}) and (\ref{KGam02})]. Here $\phibf$ denotes the order parameter with modulus $|\phibf|=\sqrt{2\rho_{0,k}}$ and arbitrary direction. For the numerical solution of the RG equations we consider dimensionless variables expressing all quantities in units of the running momentum scale $k$ so that the QCP manifests itself as a fixed point of the RG equations. The latter are solved numerically with the explicit Euler method and a discretization of the (properly adimensionalized) $\rho$ and $\wn$ variables, and an exponential regulator function $R_k(\q^2)=\alpha Z_k(0) \q^2/(e^{\q^2/k^2}-1)$ with an adjustable parameter $\alpha$ as in Ref.~\onlinecite{Rose17}.

\begin{figure} 
\centering
\includegraphics{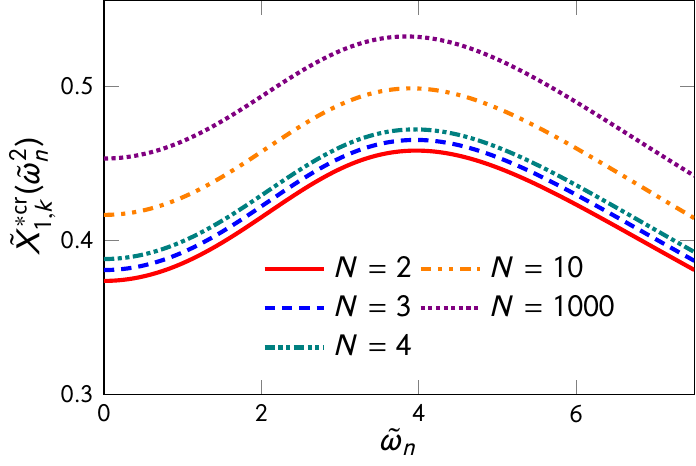}
\caption{Fixed-point function $\tilde X^{*\rm cr}_1(\twn^2)$ at the QCP for various values of $N$.}
\label{fig_X1cr} 
\centering

\includegraphics{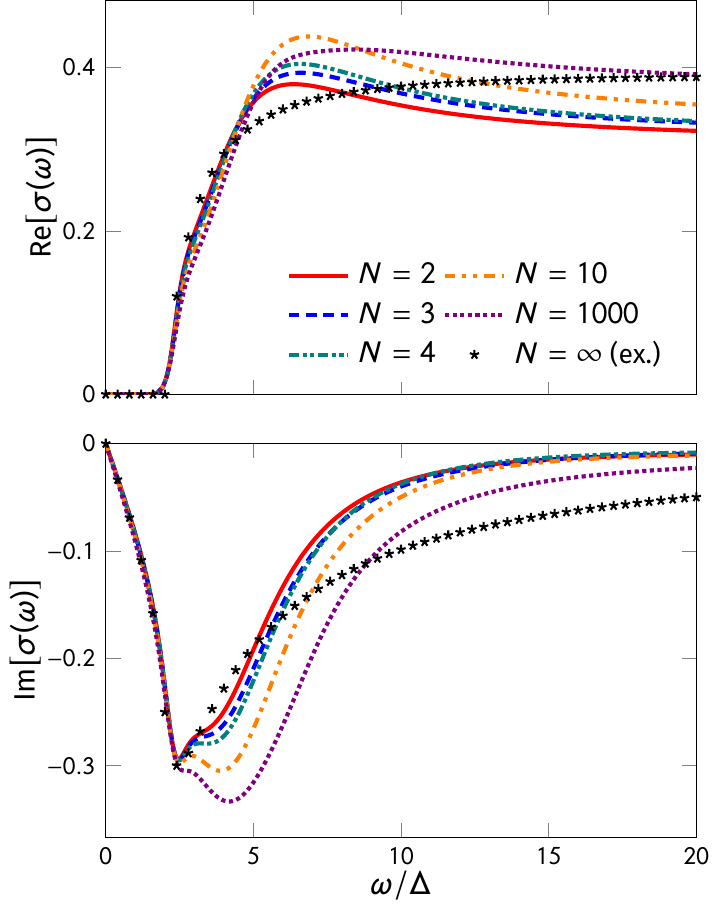}
\caption{Real and imaginary parts of the conductivity $\sigma(\w)$ in the disordered phase for various values of $N$. Black stars show the exact large-$N$ solution.}
\label{fig_sigdis} 
\end{figure}

\begin{table}
\centering
\setlength{\tabcolsep}{8pt}
\caption{Universal conductivity $\sig^*/\sigQ$ at the QCP,\cite{not6} obtained with a regulator parameter value of $\alpha=2.25$, compared to results obtained from quantum Monte Carlo simulations\cite{Witczak14,Chen14,Gazit13a,Katz14,Gazit14} (QMC) and conformal bootstrap\cite{Kos15} (CB). The exact value for $N\to \infty$ is $\pi/8\simeq 0.3927$.}
\begin{tabular}{cccc}
\hline \hline
$N$ & NPRG & QMC & CB \\
\hline
2 	& 0.3218 & 0.355-0.361 & 0.3554(6) \\
3	& 0.3285 \\
4 	& 0.3350\\ 
10	& 0.3599 \\ 
1000& 0.3927 \\ 
\hline \hline
\end{tabular}
\label{table_sigQCP}
\end{table}

\begin{figure*}[ht!]
\centering
\includegraphics[width=17.5cm]{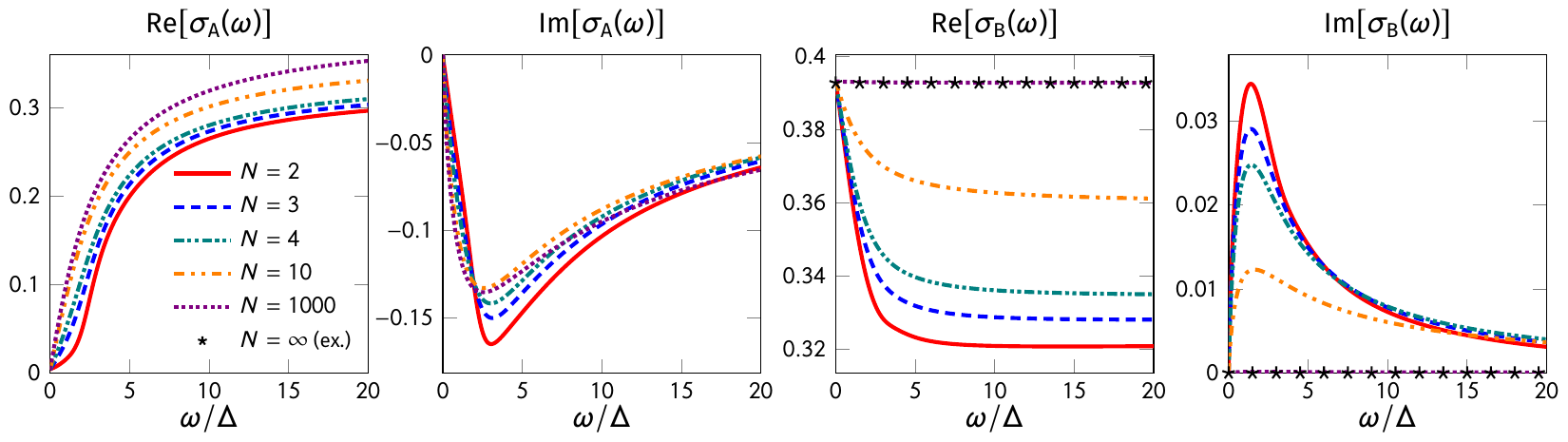}
\caption{Real and imaginary parts of the conductivity in the ordered phase for various values of $N$. Left: $\sigA(\w)$ with the superfluid contribution subtracted. Right: $\sigB(\w)$. Black stars show the exact large-$N$ solution $\sigB(\w)=\pi/8$.} 
\label{fig_sigord} 
\end{figure*}

{\bf Conductivity.}
At the QCP we expect $\sigma(\w\to 0)/\sigQ$ to take a nonzero universal value $\sig^*/\sigQ$.\cite{Fisher90,not6}
The $k$-dependent conductivity $\sig_k(i\wn)$, as a function of the Matsubara frequency $\wn$, is given by 
\beq
\sig_k(i\wn) = 2\pi\sigQ \wn X_{1,k}(\wn^2) = 2\pi\sigQ \twn \tilde X_{1,k}(\twn^2) 
\label{sigdis}
\eeq
when the order parameter vanishes ($\rho_{0,k}=0$). Here $\twn=\wn/k$ is a dimensionless frequency and $\tilde X_{1,k}(\twn^2)=k X_{1,k}(\wn^2)$ a dimensionless function of $\twn^2$. At the QCP, the function $\tilde X_{1,k}$ reaches a $k$-independent fixed-point value $\tilde X^{*\rm cr}_1$ and the conductivity takes the form $\sig_k(i\wn)=2\pi\sigQ\twn \tilde X^{*\rm cr}_1(\twn^2)$. The low-frequency universal conductivity is obtained by taking first the limit $k\to 0$ and then $\wn\to 0$, i.e. $\twn\to\infty$: $\sig^*/2\pi\sigQ=\lim_{\twn\to\infty}\twn \tilde X^{*\rm cr}_1(\twn^2)$ is thus determined by the $1/\twn$ behavior of $\tilde X^{*\rm cr}_1(\twn^2)$ at high frequencies (Fig.~\ref{fig_X1cr}). Note that this $1/\twn$ high-frequency tail, which corresponds to a $1/\wn$ divergence of $X_{1,k=0}(\wn^2)$ for $\wn\to 0$, is responsible for the breakdown of the derivative expansion of $\Gamma_k$ used in Ref.~\onlinecite{Rose17}. The value of $\sig^*$ depends weakly on the regulator, through the arbitrary parameter $\alpha$.\footnote{There is no optimal value of $\alpha$ which extremizes the value of the conductivity, i.e. $\alpha_\mathrm{opt}$ such that $d \sig^*/d \alpha|_{\alpha_\mathrm{opt}}=0$. However $\sig^*$ varies at most only by a few percents when $\alpha$ varies in the range $[1,100]$. We retain $\alpha =2.25$, which yields the best estimates for the critical exponents.} This dependence on $\alpha$ decreases as $N$ increases, and at $N=\infty$ the results do not depend on $\alpha$ and the exact value $\sig^*/\sigQ=\pi/8$ is recovered.\cite{Sachdev_book}
The universal conductivity $\sig^*$ is shown in Table~\ref{table_sigQCP} for various values of $N$. For $N=2$ we find a value in reasonable agreement with (although 10\% smaller than) results from QMC\cite{Witczak14,Chen14,Gazit13a,Katz14,Gazit14} and conformal bootstrap.\cite{Kos15}

\begin{figure}
\centering
\includegraphics{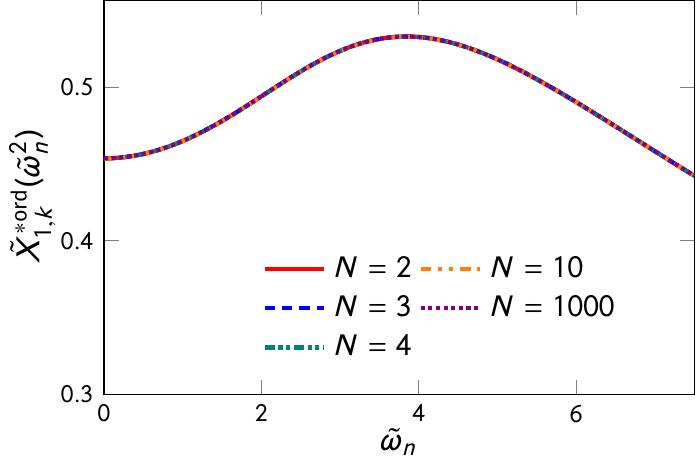}
\caption{Fixed-point function $\tilde X^{*\rm ord}_1(\twn^2)$ in the ordered phase for various values of $N$. The universal conductivity $\sigB^*$ is determined by the $1/\twn$ high-frequency tail and the collapse of the curves indicates that $\sigB^*$ is $N$-independent.}
\label{fig_X1ord} 
\end{figure}

In the disordered phase, away from the QCP, Eq.~(\ref{sigdis}) still holds since the order parameter vanishes. In Fig.~\ref{fig_sigdis} we show the real-frequency conductivity $\sig(\w)$ obtained from $\sig_{k=0}(i\wn)$ by analytical continuation using Pad\'e approximants,\cite{Vidberg77} a method which has proven to be reliable in the NPRG approach.\cite{Rose15,Rose16a,Rose17} As expected, the system is insulating. The real part of the conductivity vanishes below the two-particle excitation gap $2\Delta$ and the system behaves as a perfect capacitor for $|\w|\ll\Delta$, i.e. $\sig(\w)\simeq -i\Cdis\w$, with capacitance (per unit area) $\Cdis=2\pi\hbar\sigQ X_{1,k=0}(\wn^2=0)$. Note that for large-$N$, there is a discrepancy between the exact solution and our computation. Indeed, unlike at the QCP and in the ordered phase, in the disordered phase the  LPA$''$ does not reproduce the large $N$ solution.\footnote{This is due to the fact that the full $\rho$ dependence of the functions $Z_k$, $Y_k$, $X_{1,k}$ and $X_{2,k}$ is not taken into account.} Furthermore, the analytic continuation is made difficult by the singularity at $\omega = 2 \Delta$ so that the frequency dependence of $\sig(\w)$ above $2\Delta$ should be taken with caution.  

Let us finally discuss the two elements, $\sigA$ and $\sigB$, of the conductivity tensor in the ordered phase where the O($N$) symmetry is spontaneously broken: 
\beq
\begin{split}
\sigAk(i\wn) ={}& 2\pi\sigQ \{ \wn X_{1,k}(\wn^2) 
                 + (2 \rho_{0,k}/\hbar) \\ & \times [ Z_k(\wn^2)/\wn + \wn X_{2,k}(\wn^2) ] \}  , \\ 
\sigBk(i\wn) ={}& 2\pi\sigQ \wn X_{1,k}(\wn^2) .
\end{split}
\eeq 
At low frequencies, $\sigAk(i\wn)/2\pi\sigQ\simeq 2\rho_{0,k}Z_k(0)/\hbar\wn$ is characteristic of a superfluid system with stiffness $\rho_{s,k}=2\rho_{0,k}Z_k(0)$ (i.e. a perfect inductor with inductance $\Lord=\hbar/2\pi\sigQ\rho_s$). $\sigA(\w)$, with the superfluid contribution subtracted, is shown in Fig.~\ref{fig_sigord}. Our results seem to indicate the absence of a constant $\calO(\wn^0)$ term in agreement with the predictions of perturbation theory.\cite{Podolsky11} Furthermore we see a marked difference in the low-frequency behavior of the real part of the conductivity between the cases $N=2$ and $N\neq 2$, but our numerical results are not precise enough to resolve the low-frequency power laws (predicted\cite{Podolsky11} to be $\w$ and $\w^5$ for $N\neq 2$ and $N=2$, respectively). On the other hand we  find that $\sigB(\w)$ reaches a nonzero universal value $\sigB^*$ in the limit $\w\to 0$ (Fig.~\ref{fig_sigord}). As for the conductivity at the QCP, this universal value is determined by the $1/\twn$ high-frequency tail of the fixed-point value $\tilde X_1^{*\rm ord}(\twn^2)$ of the dimensionless function $\tilde X_{1,k}(\twn^2)$. Quite surprisingly, and contrary to $\tilde X_1^{*\rm cr}$ (Fig.~\ref{fig_X1cr}), $\tilde X_1^{*\rm ord}$ turns out to be $N$ independent: the relative change in $\tilde X_1^{*\rm ord}$ is less than $10^{-6}$ when $N$ varies (Fig.~\ref{fig_X1ord}). Noting that the obtained value $\sigB^*/\sigQ\simeq 0.3927$ is equal to the large-$N$ result\cite{Rose17,Lucas17} $\pi/8$ within numerical precision, we conjecture that $\sigB^*/\sigQ=\pi/8$ for all values of $N$. 

This result can be simply understood by noting that Goldstone bosons become effectively noninteracting in the infrared limit.\footnote{The infrared asymptotic freedom of the Goldstone bosons is explicit in the nonlinear sigma model description where the coupling constant vanishes in the low-energy limit in the ordered phase.} Using the renormalized Goldstone-boson propagator $G_{\rm T}(\p,i\wn)=[Z(\p^2+\wn^2)]^{-1}$, where $Z=\rho_s/2\rho_0$ is the field renormalization factor, and noting $Z_B$ the renormalization factor associated with the boson--gauge-field interaction for a class B generator, an elementary calculation then gives $\sigB^*/\sigQ=(Z_B/Z)^2\pi/8$. 
Gauge invariance implies that $Z_B$ and $Z$ are not independent but related by the Ward identity $Z_B=Z$, so that we finally obtain $\sigB^*/\sigQ=\pi/8$ in agreement with the NPRG result.\footnote{A similar Ward identify exists in Fermi-liquid theory, ensuring that the quasi-particle weight does not appear in physical response functions.}  

{\bf Conclusion.} We have determined the frequency-dependent zero-temperature conductivity near a relativistic $(2+1)$-dimensional QCP with an O($N$)-symmetric order parameter. Our results are obtained using the LPA$''$, an approximation of the exact RG flow equation satisfied by the effective action which respects the local gauge invariance of the theory while retaining the full momentum/frequency dependence of the vertices. Besides the frequency dependence of the conductivity both in the ordered and disordered phases, our main result is the conjecture that $\sigB(\w\to 0)/\sig_Q$ takes the superuniversal ($N$-independent) value $\sigB^*/\sigQ=\pi/8$.\cite{not6} This result could in principle be confirmed experimentally in two-dimensional quantum antiferromagnets, where both quantum criticality and the Higgs amplitude mode have been recently observed,\cite{Jain17,Souliou17} although the frequency-dependent spin conductivity has not been measured so far. A natural continuation of this work would be to extend the NPRG procedure to finite temperatures to investigate both the collisionless ($\w\gg T$) and hydrodynamic ($\w\ll T$) regimes, the latter being inaccesible at zero temperature.

{\bf Acknowledgment}
We thank N. Defenu for pointing out Refs.~\onlinecite{Casetti11,Nerattini14} and P. Kopietz for correspondence.


\begin{thebibliography}{33}
\bibitem [{not()}]{not6}%
  \BibitemOpen
  \href@noop {} {}\bibinfo {note} {We consider only zero-temperature coherent
  transport where the DC conductivity is defined by the limit $T=0$ and $\w\to
  0$ or, equivalently, $T,\w\to 0$ with $\w/T\to\infty$.}\BibitemShut {Stop}%
\end{thebibliography}

%

\end{document}